Т.В. Ганджа, С.А. Панов

# Задачи и архитектура подсистемы документирования исследований в среде многоуровневого моделирования МАРС

Рассматривается система автоматизации документирования исследований, предназначенная для построения отчетов о проведенных компьютерных исследованиях сложных технических объектов и систем в среде многоуровневого моделирования МАРС. Сформулированы назначение, задачи и алгоритм работы системы документирования исследований, рассмотрены типы и структура документов, а также приведен пример ее практического использования.
**Ключевые слова:** система, документ, отчет, исследование, моделирование.

**1. Назначение и задачи системы документирования исследований.** На сегодняшний момент в сфере образования и науки актуальной является проблема автоматизации документирования результатов анализа компьютерных моделей сложных технических устройств и систем. В основном это относится к техническим дисциплинам, в которых необходимо проведение практических занятий и лабораторных исследований (например, проведение эксперимента). Разработка системы автоматизации документирования исследований (САДИ) позволит решить проблему генерации отчетных форм, в которых содержатся результаты выполняемых исследований.

В Томском государственном университете систем управления и радиоэлектроники разрабатывается и развивается среда многоуровневого моделирования МАРС (СММ МАРС) [1]. Основанная на методе компонентных цепей (МКЦ) [2], она позволяет представить физически неоднородный технический объект (электрическая схема, кинематической преобразователь, термодинамическая система и т.д.) в виде компонентной цепи и произвести его моделирование в статическом и динамическом режимах. В настоящее время СММ МАРС нашла широкое применение в учебном процессе при проведении исследований в рамках практических и лабораторных работ, а также при выполнении курсовых и выпускных работ.

С целью автоматизации формирования отчетов о проводимых исследованиях в рамках СММ МАРС разрабатывается подсистема документирования исследований. Она должна существенно облегчить задачу оформления результатов анализа проведенных лабораторных работ по таким техническим дисциплинам, как ТОЭ, ТАУ, электроника, за счет автоматизации процесса документирования. САДИ выполняет следующие задачи: формирование отчета о проведенных исследованиях (с использованием библиотеки моделей компонентов, предназначенных для формирования отчетов), отправка сформированных отчетов в «Microsoft Office Word» и в систему управления проектами (СУП). Данный отчет содержит промежуточные и итоговые результаты моделирования, представленные в виде числовых значений, схем, графиков, таблиц и других единиц. Итоговый отчет представляется в виде документа и может быть:

- открыт в текстовом редакторе «Microsoft Office Word» для окончательного редактирования и распечатки;
- сохранен в системе управления проектами (СУП);
- проверен преподавателем без непосредственного участия студента.

**2. Место системы документирования исследований в архитектуре системы многоуровневого моделирования «МАРС».** Основной задачей среды моделирования МАРС, основанной на методе компонентных цепей, является моделирование сложных технических объектов и систем (СТУС) во временной и частотной области.

Для решения этой задачи применяется многоуровневая архитектура среды моделирования МАРС (рис. 1), состоящая из трех уровней обработки информации, выраженных на соответствующих слоях редактора: визуальном, логическом и схемном.

На схемном слое модель исследуемого объекта представляется в виде компонентной цепи как совокупности взаимосвязанных между собой компонентов. Каждому компоненту схемного слоя ставится в соответствие вычислительная модель, реализующая этот компонент в рамках универсального вычислительного ядра [3].

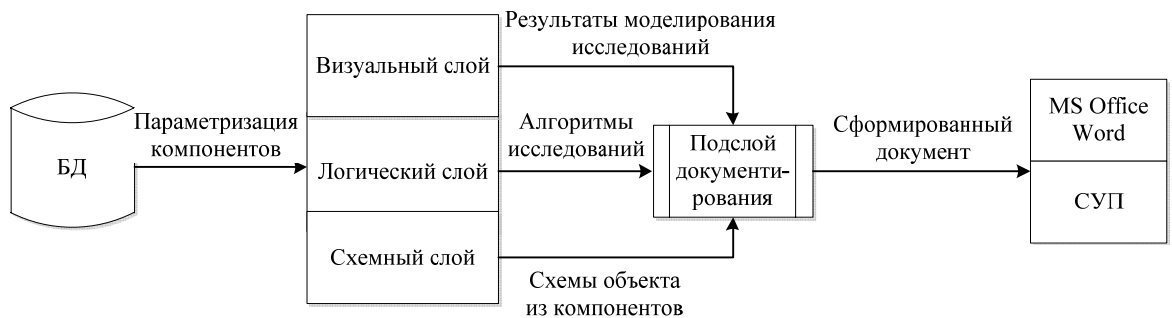

Рис. 1. Место САДИ в архитектуре СММ «МАРС»

Моделирование компонентной цепи, полученной в результате соединения и параметризации всех компонентов, производится путем автоматического формирования и расчета системы алгебраических и дифференциальных уравнений в статическом или динамическом режиме. В компонентную цепь включаются измерительные компоненты, предназначенные для измерения и передачи результатов на средства визуализации, такие как цифровые табло, графики, таблицы и т.д. Используя компонентный подход, переданные в них результаты могут быть обработаны компонентами-блоками обработки результатов моделирования, с целью получения интегральных значений, называемых параметрами-функционалами.

На визуальном слое формируется лицевая панель виртуального измерительного прибора или стенда системы управления исследуемой моделью технического объекта. На логическом слое данные компоненты имеют отображение со связью для получения подлежащих визуализации результатов, а также для передачи в алгоритмы управления пользовательской информации.

На логическом слое формируются алгоритмы обработки результатов измерения и моделирования в виде одной или нескольких цепочек компонентов. Информация между ними осуществляется с помощью механизма обмена сообщениями. Именно на логическом слое (в виде подслоя) реализуется САДИ. Представление объекта и алгоритмов его моделирования в формате компонентных цепей требуют реализации САДИ в виде специализированных компонентов. На данном слое располагаются различные компоненты, основанные на интерактивной математической панели. Она имеет в своем составе редактор математических выражений, позволяющий вводить и редактировать математические выражения, которые могут быть предварительными расчетами параметров компонентов, блоками обработки результатов моделирования либо целевыми функциями для решения задач оптимизации и параметрического синтеза.

САДИ должна позволять добавлять в формируемый документ:
– схемы исследуемого объекта, представленные на схемном слое редактора СММ МАРС;
– результаты моделирования, представленные статическими значениями, графиками, таблицами и диаграммами на визуальном слое редактора СММ МАРС;
– отдельные формулы, набранные в редакторе математических выражений интерактивной математической панели, алгоритмы решения технических задач, представленные как блок-схемами, набранными из компонентов, так и последовательностью математических выражений, на логическом слое редактора СММ МАРС.

Помимо этого, добавляемые элементы должны сопровождаться текстовыми комментариями и названиями определенного вида. Сформированный документ должен открываться в широко использующихся текстовых редакторах, таких как «Microsoft Office Word», «Open Office» и др.

Такая система позволит сформированный один раз шаблон документа использовать его же и для исследования ряда подобных объектов. При изменении параметров и структуры исследуемого объекта, а также алгоритма его исследования результаты моделирования будут автоматически изменяться в формируемом документе.

Среда многоуровневого моделирования МАРС, в рамках которой предложено разработать систему документирования исследований, основывается на методе компонентных цепей [2]. Данный метод позволяет представить в виде компонентной цепи модель исследуемого объекта, алгоритм ее исследования, а также систему визуализации и обработки результатов экспериментов. Логичным является представление в виде компонентной цепи процесса формирования документа, содержащего результаты исследования.

**3. Использование метода компонентных цепей для формирования документов в СММ МАРС.** В рамках СММ МАРС реализованы методы математического и имитационного моделирования. Математическое моделирование используется для компьютерного исследования технических объектов. Оно предполагает формирование и расчет системы уравнений объекта в статическом и динамическом (во временной и частотной области) режимах. Имитационные методы применяются для обработки результатов моделирования и моделирования алгоритмов управления техническими объектами. Они предусматривают передачу информации между связанными компонентами. Эти принципы в рамках СММ МАРС лежат в основе разработки компонентов для автоматизированного формирования отчетов о проведении экспериментов. Для этих целей в рамках расширения метода компонентных цепей для генерации отчетных форм реализуются следующие компоненты.

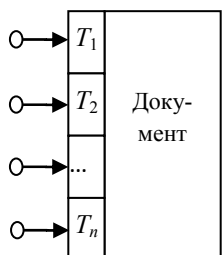

Рис. 2. Компонент «Документ»

**Обобщенный компонент «Документ»** (рис. 2) позволяет автоматически сформировать отчет о выполняемых в рамках СММ МАРС исследованиях. Для этого к нему подключается множество компонентов-поставщиков данных различных типов. Каждой связи данного компонента соответствует тег с именем $T_i$ ($i = 1, 2, …, n$). Не допускается наличие двух тегов с одинаковым именем для двух и более компонентов, т.е. $T_i \ne T_j$. По двойному щелчку мыши по изображению данного компонента в редакторе схем открывается редактор, в котором пользователь формирует шаблон документа и тегами указывает те места, куда будут вставляться данные, поступающие по связям.

В результате комплексного моделирования компонентной цепи, содержащей компоненты технического объекта, блоков обработки результатов и описываемые компоненты, компонент «Документ» будет содержать отчет, в который будут включены результаты моделирования технического объекта.

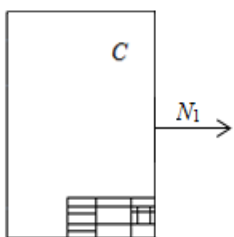

Рис. 3. Компонент «Схема»

**Компонент «Схема»** (рис. 3) применяется для добавления в отчет схем, созданных на схемном слое многослойного редактора СММ МАРС. При добавлении новой схемы в отчет требуется выделить нужную схему с помощью мыши, растянув прямоугольник, и в открывшемся диалоговом окне поименовать ее. Все поименованные схемы будут иметь следующие свойства: имя, координаты границ. Свойство «Имя» служит для выбора нужной схемы из списка доступных и для ее добавления в определенное место отчета. Свойство «координаты» задается автоматически и служит для хранения координат схемы.

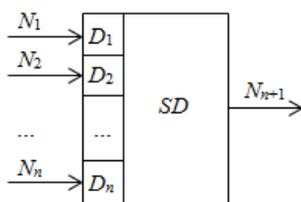

Рис. 4. Компонент «Диаграмма»

**Компонент «Диаграмма»** (рис. 4) позволяет добавить в отчет динамические и статические диаграммы. Динамические диаграммы служат для отображения графиков переходных процессов, частотных характеристик, вольт-амперных характеристик, параметрических характеристик и векторных диаграмм. Динамические диаграммы добавляются в отчет по запросу пользователя и могут менять свои значения в зависимости от входных данных, которые пользователь может менять динамически. Статические диаграммы применяются для отображения графиков со статическими значениями и строятся в отдельном окне СММ МАРС. При добавлении такой диаграммы в отчет необходимо указать в ее свойствах имя нужной диаграммы.

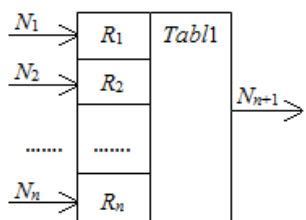

Рис. 5. Компонент «Формирователь таблицы»

**Компонент «Формирователь таблицы»** (рис. 5) применяется для добавления в отчет таблиц. У компонента «Формирователь таблицы» имеются следующие свойства: имя, количество строк и столбцов, данные ячеек. Данные ячеек таблицы вводятся пользователем самостоятельно либо поступают в ячейку автоматически с компонентов-блоков обработки результатов моделирования и имеют различные типы данных (текст, число, график, рисунок и т.д.).

С целью добавления сформированных элементов (схем, графиков и диаграмм, таблиц с результатами исследования) в отчет с возможностью изменения данных, получаемых при моделировании, соответствующие компоненты своим единственным выходным узлом, расположенным справа от изображения компонента, присоединяются к компоненту «Документ» (см. рис. 2).

**4. Алгоритм работы системы документирования исследований.** СММ МАРС используется в процессе обучения студентов для выполнения различных лабораторных и научных исследований по техническим дисциплинам (ТАУ, ТОЭ, электроника и т.д.). Для СММ МАРС разрабатывается модуль САДИ, который позволит студентам автоматически формировать отчеты о проделанной работе. Эти отчеты для непосредственного оценивания выполненных работ, для их передачи на другие компьютеры посредством Интернет или системы управления проектами.

Рассмотрим алгоритм работы САДИ на примере выполнения студентом лабораторной работы. Данный алгоритм включает в себя:

1. **Аутентификацию студента.** Регистрация нужна для того, чтобы контролировать работу студентов.
2. **Формирование компонентной схемы.** Формирование схем в СММ МАРС выполняется из доступных компонентов.
3. **Параметризацию схемы.** Параметризация осуществляется вводом нужных значений в свойства компонентов.
4. **Анализ.** После окончания формирования компонентной схемы, студент запускает процесс анализа, в результате выполняется построение графика и расчет статических значений. При необходимости производится коррекция параметров схемы.
5. **Обработку результатов.** Выполняется с целью определения значений параметров-функционалов, характеризующих временную или частотную характеристику.
6. **Оформление отчета.** С помощью САДИ можно автоматически формировать отчет, который сохраняется в виде документа. После того как отчет сохранен, студент может самостоятельно загрузить его в СУП.

СУП – Web-приложение, разработанное специально для того, чтобы управлять различными типами научно-исследовательских работ студентов (НИРС), а также осуществляющее координацию между исполнителями и руководителем НИРС [5].

Преподаватель может выполнять следующие действия с документом:

1. Если документ сохранен в виде файла: открыть его в текстовом редакторе «Microsoft Office Word» для редактирования, проверки или распечатки.
2. Если документ загружен студентом в СУП:

- Просмотр документа. Существует два режима просмотра отчетов в СУП: непосредственно в самой СУП и в виде, готовом для вывода на печать.
- Редактирование документа. Все элементы документа можно в любое время отредактировать.
- Удаление документа. Преподаватель может удалить документ.
- Архивирование документа. Преподаватель может отправить документ в архив для последующего хранения.

Студент может выполнять следующие действия в СУП: зарегистрироваться в качестве студента; загружать различные документы в систему; просматривать, редактировать или удалять документы, к которым разрешен доступ; осуществлять координацию с руководителем НИР; обмениваться сообщениями с другими студентами.

**5. Структура и типы документов в системе документирования исследований.** Основным объектом управления в САДИ является документ. Документ может содержать различные типы данных: числа, структурированный текст, графику, математические выражения и формулы, организационные диаграммы и т.д.

Типичный документ содержит следующую структуру: титульный лист, аннотациию, реферат, содержание, введение, основную часть, заключение, список использованной литературы и приложения. Необходимо отметить, что в некоторых типах документов могут отсутствовать некоторые элементы, такие как аннотация, реферат и т.д.

К студенческим работам, имеющим представленную структуру, относятся: расчетно-графическая работа, курсовая работа, курсовой проект, дипломная работа и дипломный проект.

Основной задачей САДИ является автоматизированное формирование документов указанных типов, содержащих результаты выполненных в СММ МАРС исследований.

Рассмотрим пример документа на примере отчета по лабораторной работе по ТОЭ. Отчет должен содержать:

1. Исследуемую схему, расчетные формулы основных величин:
2. Таблицу рассчитанных величин.
3. Графики электрических характеристик, указанных в работе.
4. Ответы на контрольные вопросы.

Для формирования отчета необходимо использовать различные компоненты из библиотеки моделей компонентов (БМК).

В лабораторных работах по ТОЭ обычно используются следующие компоненты: документ (см. рис. 2), схема (см. рис. 3), диаграмма (см. рис. 4), формирователь таблицы (см. рис. 5).

Для автоматического формирования отчета создается шаблон отчета, в котором с помощью тегов указываются места, куда автоматически заносятся результаты моделирования.

**Заключение.** При компьютерном исследовании различных технических объектов и (или) их виртуальных аналогов важным вопросом являются документирование, сохранение и обработка результатов моделирования. В данной работе предложена и исследована система автоматизации документирования исследований, представляющая собой модуль к многоуровневой среде моделирования МАРС.

Система автоматизации документирования исследований позволит оперативно обмениваться по Интернету документами, содержащими результаты исследований, проведенных в рамках среды многоуровневого моделирования МАРС, между группой людей, работающих на большом расстоянии друг от друга.

___

**Ганджа Тарас Викторович**
Канд. техн. наук, доцент ВКИЭМ ТУСУРа
Тел.: (382-2) 41-39-15
Эл. почта: gandgatv@gmail.com

**Панов Сергей Аркадьевич**
Аспирант ВКИЭМ ТУСУРа
Тел.: 8-961-885-24-00
Эл. почта: spytech3000@gmail.com

Gandzha T.V., Panov S.A.
**Tasks and architecture of documentation subsystem in multi-level modeling environment MARS**

The article describes the automated documentation system designed to generate reports on research conducted by computer complex technical objects and systems in multi-level modeling environment «MARS». We defined the purposes, tasks and abilities of documentation system and examined the types and structure of documents, and gave an example of its practical use.
**Keywords**: systems, documents, reports, research, modeling.

___